\documentstyle[prl,aps,psfig]{revtex}
\begin{document}
\draft

\twocolumn[\hsize\textwidth\columnwidth\hsize\csname
@twocolumnfalse\endcsname

\widetext
\title{ Anomalous finite size spectrum  in the $S=1/2$ 
 two dimensional Heisenberg model }
\author {  Catia Lavalle$^{2}$, Sandro Sorella$^{1,2}$ 
and 
Alberto Parola$^{1,3}$}
\address{ 
$^1$ Istituto Nazionale di Fisica della Materia \\
$^{2}$  International School for Advanced Studies,
Via Beirut 2-4, I--34013 Trieste, Italy \\
$^3$ Istituto di Scienze Fisiche, Universit\'a di Milano, Via Lucini 3 Como,
 Italy } 

\maketitle
\begin{abstract}
We  study  the low energy spectrum of 
the nearest neighbor Heisenberg model  on a square lattice 
as  a function of the total spin $S$.
By quantum Monte Carlo simulation we compute this spectrum for the $s=1/2$, 
$s=1$ and $s=3/2$ Heisenberg models.
We conclude that the nonlinear $\sigma$ model
prediction  for the low energy spectrum  is always verified    for large enough
system  size. However the crossover to the  correct
scaling regime is particularly slow just for the $s=1/2$ Heisenberg model.
The possibility to detect  this unexpected
anomaly  with finite temperature experiments on $s=1/2$ isotropic quantum
antiferromagnets   is  also discussed.

\end{abstract}

]

\narrowtext

The square lattice Heisenberg model (HM) has attracted much attention in 
recent years because of 
its connection with the antiferromagnetic properties of the undoped
stoichiometric compounds of High-Tc superconductors\cite{Pines}. The model 
Hamiltonian in general dimension $d$, reads:
\begin{equation} \label{hamiltoniana}
H= J\,\sum_{<i,j>}   \vec S_i \cdot \vec S_j 
\end{equation}
where the symbol $<i,j>$ indicates nearest neighbor summations, 
 the index $i$ labels the positions $R_i$ of the 
$N=L^d $ sites on a hypercubic lattice and
the quantum spin operators satisfy $( \vec S_i )^2 = s (s+1)$.
What makes particularly difficult any analytical treatment of this model
is the fact that the antiferromagnetic
order parameter $\vec m ={1\over N} \sum_i e^{iQ\cdot R_i} \vec S_i $  
(with $Q=(\pi,\pi...)$) does not commute either with the
Hamiltonian or with the total spin $\vec S= \sum_i \vec S_i$.
Whenever long range order is present in the thermodynamic limit, a huge 
degeneracy of the energy spectrum $E(S)$ as a function of the total 
spin  $S$ is implied by the mentioned non commutativity. Hence $E(S)$,
referenced to the singlet $S=0$ ground state energy, 
is predicted to behave as the spectrum of a free quantum
rotator as long as $S \ll \sqrt{N}$ \cite{Anderson,Ziman,Fisher}:
\begin{equation} \label{top}
E(S) = S ( S+1) / (2 I N) 
\end{equation}
where $I$ is known as the {\it momentum of inertia} per site. 

This equation resembles the definition of the spin susceptibility $\chi$
which is however obtained by taking {\it first} the infinite volume limit of 
the energy per site $e(m)=E(S)/N$ at fixed magnetization $m=S/N$ and then 
letting $m\to 0$: $ e(m)= m^2/(2 \chi) + o(m^2) $. 
An identification between $I$ and $\chi$ is possible only if 
the excitation spectrum smoothly connects the low energy portion, which 
corresponds to total spin $S\sim O(1)$, with the regime of 
macroscopic spin excitations: $S \sim m N$ (with $m << 1 $).
This is a highly non trivial statement which is actually verified by 
the underlying low energy model of the quantum antiferromagnet (QAF), known 
as the non linear $\sigma$ model (NL$\sigma$M) \cite{Chakravarty}.

For the HM in one dimension the accepted low energy model is instead
the Luttinger
liquid with $K_\sigma=1/2$ and the prediction in this case is that the energy
spectrum as a function of the total spin behaves as $E(S) = S^2/(2 \chi N)$  
where $\chi$ coincides with the spin susceptibility \cite{Hald}. 
Note the strong analogy of the spectrum in one and higher dimensions,
though in the former case no true long range order sets in.   

It is particularly important to verify that,
also in presence of long range order, the low energy spectrum of the 
microscopic Hamiltonian conforms to the prediction of the low energy effective 
model. This is the main problem addressed in this paper where 
we show the results of high accuracy computations of the 
the energy spectrum $E(S)$ in few two dimensional non frustrated
spin models. To this purpose  we use an improved version of the known 
lattice  Green's  function Monte Carlo (GFMC)
technique introduced some years ago by Trivedi and Ceperley\cite{Ceperley}.
Our method allows to control any form of systematic error in a
rigorous and simple way \cite{Sorella}. For large lattice size, the energy 
difference between spin subspaces becomes extremely small.  
Nevertheless our technique allows a good resolution of the spectrum 
$E(S)$ up to a $16\times 16$ lattice with a reasonable computational effort. 
The method we use  allows to eliminate exactly any source of systematic
error which affects previous GFMC calculations \cite{Runge}.
For large size the results by Runge are also affected by a non negligible 
population control error which is exactly  eliminated in this GFMC 
scheme \cite{Sorella}. Our results for the ground state energy are instead 
consistent with those shown in Ref. \cite{Sandvik}, obtained with a 
completely different QMC method \cite{byline}.  
 
If the quantum top law (\ref{top}) is verified, we can calculate 
the unknown quantity $1/2 \chi$ on a finite lattice by inserting the
computed excitation energies at different spin $S$ in the equation
\begin{equation} \label{chi_S} 
[2\chi_S]^{-1} = N \,E(S)\,[S (S+1)]^{-1}
\end{equation} 

Clearly, $1/\chi_S$ should approach the physical inverse
susceptibility  for infinite size and for any spin excitation $S<<N$ 
provided  the quantum top law is verified. 

\begin{figure}
\centerline{\psfig{figure=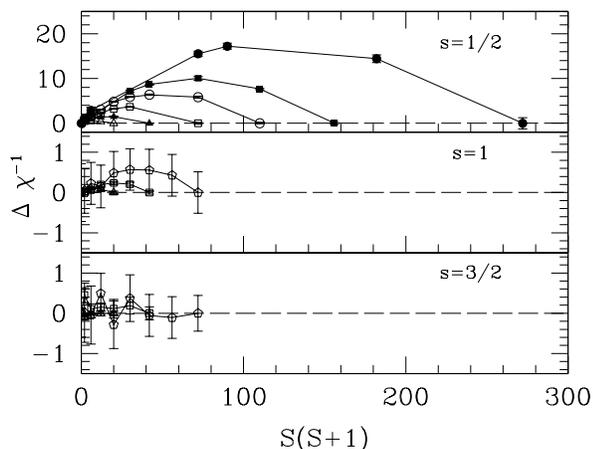,height=6cm}}
\caption {\baselineskip .185in\label{plot_anomaly_iso}
$\Delta \chi^{-1}$ for 2D HM of
spin $s=1/2$ (data up to N=$256$),
$s=1$ (data up to N=$100$) and $s=3/2$ (data up to N=$64$).
Lines are guides to the eye.}
\end{figure}

It is then instructive to 
study the deviations from Eq. (\ref{top}) through the quantity: 
\begin{equation} 
\Delta \chi^{-1} = \left[ {1 \over 2 \chi_S } - {1 \over 2 \chi_L} \right] S
(S+1) 
\end{equation}
which is zero for $S=0$ and $S=L<< N$, by definition and should 
vanish for any fixed $S$ in the $N \to \infty$ limit.
A finite slope in the numerical data, 
as clearly shown by Fig.~\protect\ref{plot_anomaly_iso}\protect\ for $s=1/2$, 
would imply $1/ 2 I > 1/ 2\chi_L$, with a violation of the quantum top law. 
A similar and even more pronounced, anomaly was also reported in 
Ref.~\cite{Parola} to support the existence of a spin liquid phase 
for the spatially anisotropic HM with strong anisotropy. 
Fig.~\protect\ref{plot_anomaly_iso}\protect\ provides
clear evidence that this anomaly in the spectrum surprisingly persists 
in the isotropic 2D HM for $s=1/2$ HM while is absent, within 
statistical errors, in the higher spin models shown in the same figure.
This apparently contrasts with the accepted picture stating that the 
low energy properties of the $s=1/2$ HM can be fully described 
by the NL$\sigma$M.

How can we reconcile this numerical evidence with 
the  established theories of the 2D $s=1/2$ quantum antiferromagnets? 
Does this anomaly have observable consequence in the low energy behavior of
the 2D HM ? 
In order to answer these questions it is important to have a
better  control of the finite size corrections, because the data presented 
in this paper are close to the limits of the available computers and
considerably larger size cannot be handled with the required accuracy. 

In the following, we will make extensive use of 
spin wave theory\cite{Anderson,Oguchi} (SWT) which is known to be 
extremely accurate for the HM. For instance, the recent quantum Monte Carlo 
(QMC) calculation of the order parameter gives $m \simeq 0.31$
\cite{Sorella,Runge,Sandvik},
quite close to the linear spin wave (SW) estimate $m\simeq 0.3$.
Moreover SWT, which is based on a systematic  expansion in $1/s$, 
can be generalized without basic difficulties
to finite systems, making possible a direct comparison of the 
numerical results with the spin-wave predictions \cite{Zhong}.

In order to study the spin excitation spectrum of the model as a function of
the total spin,  we  add to the HM Hamiltonian $H$ a magnetic field
along the $z$ direction: $H_h = H -h s \sum S^z_i$. The magnetic field 
$h$ can be chosen to stabilize the desired spin excitation of spin $S$. 
In order to perform a systematic SW calculation for large $s$ we have
scaled the magnetic field by $s$. In this way, when $s \to \infty$ 
the magnetic energy due to the external field $h s  \sum S^z_i $ is of the 
same order $\sim s^2 $ as the average energy $E(S) = <H>$.   
The classical solution  for $s\to \infty$ at fixed $h$ is simply 
obtained by canting the spins of an angle $\theta$  
in the direction of the field (with $ \sin \theta = { h \over 4 d} $ ).
By use of the standard Holstein-Primakoff representation of 
the spin operators in terms of bosons and by expanding the resulting
Hamiltonian in powers of $1/s$, we can compute the fluctuations
over the classical solution and obtain the linear SW estimate of the
energy of $H_h$\cite{Zhong,Franjic}:
\begin{eqnarray} \label{lsw}
E(h)&=& -J   s^2 N (d - h^2/ 8 d ) -J d s (   N -{\textstyle \sum_k}  
\epsilon_k ) \\
\epsilon_k &=&  {\textstyle{1\over 4d}}\,
\left[ 16 d^2\,(1 -\gamma_k^2)-2 h^2 \gamma_k ( 1 -\gamma_k ) \right ]^{1/2}
\nonumber
\end{eqnarray}
with $\gamma_k~=(\cos k_x + \cos k_y  + \cdots)/d $.
By expanding $E(h)$ about $h=0$, we get 
$E(h) \sim  -{1\over 2} \chi^{SW} (s h)^2 $ which immediately gives
the well known linear SW correction to the classical spin
susceptibility $\chi_0~=~1 /4dJ $: 
$ \chi_{SW}/\chi_0~=~1-0.55115/2s $\cite{Oguchi}.
This  calculation can be extended to the next leading order and gives
the second order  correction to $\chi$:
\begin{equation} \label{second}
Z_\chi = \chi/\chi_0 = 1 -0.55115 /2 s  +0.0403 /(2 s)^2 \quad .
\end{equation} 
The above expansion seems to converge rapidly also for $s=1/2$.
This result differs from the contradicting 
second order SW calculations present in
the literature \cite{Canali}.  Our SW expansion,
however, can be in principle different due to the scaling of  
the magnetic field by $s$, which considerably simplifies the $1/s$ 
expansion of $\chi$ but has not been used before \cite{Oguchi,Canali}.  

\begin{figure}
\centerline{\psfig{figure=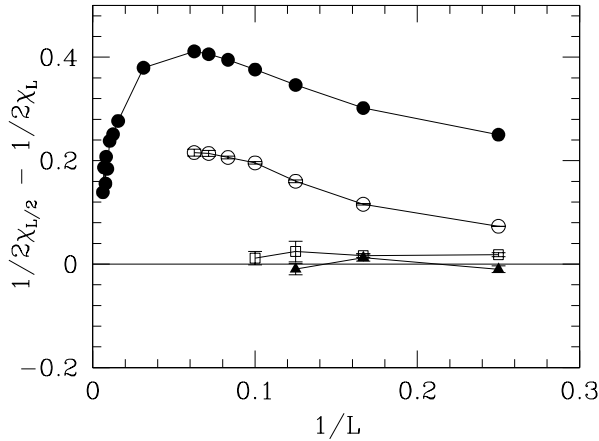,height=6cm}}
\caption {\baselineskip .185in\label{scal}
Size dependence of $1/2 \chi_S$ as defined in (\protect\ref{chi_S})
for the 2D HM. Upper curve: linear spin wave.
QMC data refers to $s=1/2$ (circles)
$s=1$ (squares) and $s=3/2$ (triangles). Lines are guides to the eye.}
\end{figure}

The energy spectrum $E(S)$ of the $s=1/2$ HM as a function of the total spin 
$S$ can be evaluated within linear SWT in two equivalent ways. We can either 
perform  the  Legendre transform $E(S)=E(h)+h s S$ 
of Eq. (\ref{lsw}) at fixed $s=1/2$ (referred to as SWH),
or we can scale the spin $S =\mu s$ with fixed $\mu$ evaluating the Legendre
transformed $E(\mu s )$  perturbatively  in $1/s$ before setting  $s=1/2$
(referred to as SWM). 
Of course these two methods are equivalent within a $1/s$ expansion
and the differences only arise because we truncate SW expansion at a 
finite order in $1/s$.
The apparent deviations from the quantum top law shown in 
Fig.~\protect\ref{plot_anomaly_iso}\protect\ 
can be understood on the basis of the aforementioned linear SW analysis. 
To this purpose, in Fig.~\protect\ref{scal}\protect\ we compare 
the SWH prediction for  $1/2 \chi_L - 1/2 \chi_{L/2}$ with the QMC data.

We see that both QMC and SWH results, although quantitatively different,
show the same dependence on lattice size with an apparent plateau
at about $L\sim 10$ which would suggest a breakdown of the
quantum top law. While QMC data stop because it is not possible 
to obtain high quality data at larger size, the SW results 
show an abrupt change of behavior with a clear convergence towards the
expected result. The correct $1/L$ scaling is however reached
only for $L > 32$.
\begin{figure}
\centerline{\psfig{figure=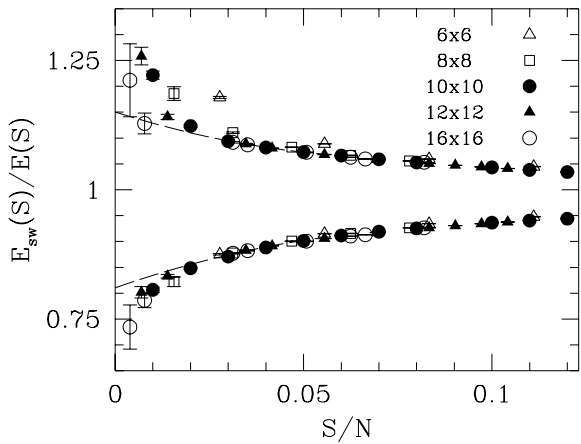,height=6cm}}
\caption {\baselineskip .185in\label{ztot}
Ratio of the QMC spectrum with linear
SWM (lower curve) and  SWH (upper curve) for several system sizes.
The curves are least squares fits of the data.  }
\end{figure}

The similarity between QMC and SWH, where both available, suggests
that the non monotonic behavior is a genuine feature of the model.
In order to provide support for this scenario, we considerably reduced 
the anomalously large finite size corrections present in 
Fig.~\protect\ref{scal}\protect\ by taking the ratio between 
the energy spectrum evaluated via QMC and SWT (using both the SWM 
and SWH methods). As shown in Fig.~\protect\ref{ztot}\protect\  
we get a remarkable collapse of all the QMC data for sufficiently large sizes
($L\sim 10 \div 16 $) onto a single curve, even at small total spin $S$. 
This is a direct confirmation that the anomalies detected in 
the spectrum of the $s=1/2$ HM are quite similar to those   
present in linear SWT, and eventually disappear only for 
very large size.  We stress that this unexpected behavior 
is visible only in the $s=1/2$ HM and is strongly reduced for 
higher $s$ (but also in spin-anisotropic systems like the $XY$ model).
This suggests that quantum fluctuations, enhanced by small values of $s$,
deeply affect the properties of the model when the spontaneous
breaking of a non Abelian symmetry occurs. 
\begin{figure}
\centerline{\psfig{figure=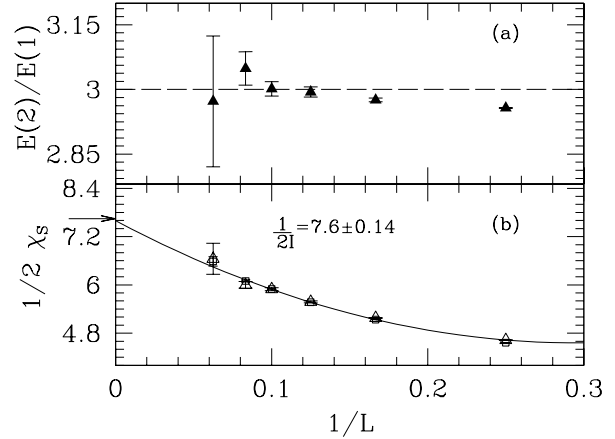,height=6cm}}
\caption {\baselineskip .185in\label{sO1}
(a) Ratio between 
the S=2 and S=1 excitation energies. (b): Size dependence of $1/2 \chi_S$ 
(\protect\ref{chi_S}) for S=1 (triangles) and S=2 (squares).  
The arrow shows the expected $1/2\chi$ 
(see Fig.~\protect\ref{ztot}\protect). The continuous line 
is a parabolic extrapolation ($1/2I=1/2\chi_S$ for $L\to\infty$) for $S=2$ and 
$6\le L\le 16$.}
\end{figure}
We also notice that 
at small spin excitations the size effects are not so 
dramatic as shown in Fig.~\protect\ref{sO1}\protect(a) where 
the QMC ratio $E(S=2)/E(S=1)$ is seen to smoothly approach the value $3$ 
predicted by the quantum top law (dashed line).  
The possibility to control the anomalously large finite size effects 
allows for an accurate numerical estimate of the spin susceptibility even 
in the $s=1/2$ HM. In fact, the susceptibility can be directly obtained by 
extrapolating to zero magnetization $m=S/N$ the QMC results shown in 
Fig.~\protect\ref{ztot}\protect. However, only data 
representative of the thermodynamic limit of the energy per site 
$e(m)=E(S)/N$ must be included in the extrapolation procedure.
As shown in Fig.~\protect\ref{ztot}\protect\ the QMC ratios to both SWH and SWM
predictions can be nicely extrapolated to $m=0$ if we exclude 
points with $m < 0.03$. On the other hand, the thermodynamic limit 
of the SW results can be obtained analytically:
$e(m)=m^2 [2 \chi_0 (1-0.55115/2 s )]^{-1}  $ and  
$e(m) =m^2  [2\chi_0]^{-1} ( 1+0.55115 /2 s) $  for the SWH  and 
SWM estimate respectively. Both methods consistently give 
the following result for the spin susceptibility of the $s=1/2$ HM:
\begin{equation} 
Z_\chi =\chi/\chi_0 =0.523 \pm 0.005 
\end{equation}
Our value for $Z_\chi$ is also consistent, 
within error bars, with the independent 
calculation of $I$  shown in Fig.~\protect\ref{sO1}\protect (b) and is
not far from the second order SW value (\ref{second}). This 
results slightly differs from series expansion  
Ref.~\cite{Singh} 
but much more from  previous finite size scaling analysis  based 
on  QMC data on similar lattice sizes \cite{Runge,Sandvik}. 
\begin{figure}
\centerline{\psfig{figure=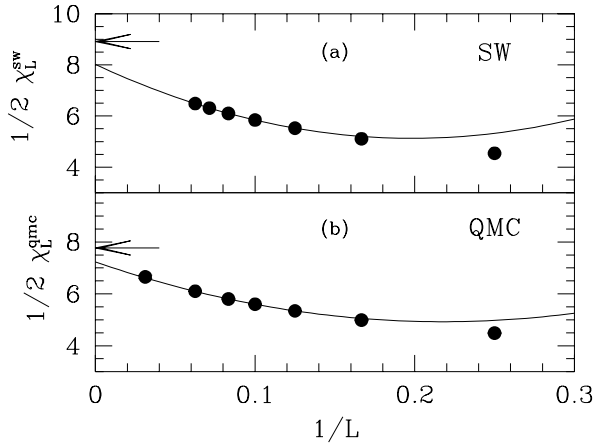,height=6cm}}
\caption {\baselineskip .185in\label{chi}
Same as Fig.~\protect\ref{sO1}\protect (b) for the spin excitation $S=L$. 
Curves are parabolic extrapolations. 
The arrows are the expected infinite size limits.}
\end{figure}
The latter  discrepancy can be easily understood on the basis of the 
strong finite 
size anomaly in the spectrum of the $s=1/2$ HM we have discussed.
In fact, as shown in Fig.~\protect\ref{chi}\protect (a) if we
extrapolate to the thermodynamic limit the known SW spectrum 
in lattices with $L < 16$ by use of the correct asymptotic scaling
$\chi_L = \chi_\infty + a /L +b/L^2 $ \cite{Fisher,Ziman}, we get 
a poor estimate of the exact $L\to \infty$ limit of $\chi_{SW}$. In fact,
SWT suggests that the correct scaling is valid only for extremely 
large sizes ($L>64$) while it is not accurate for the lattice sizes 
where QMC simulations are available (see Fig.~\protect\ref{chi}\protect (b)). 

A further evidence that the data shown in Fig.~\protect\ref{ztot}\protect\ 
are very weakly size dependent comes from the consistency 
with the NL$\sigma$M next leading $m\to 0$  prediction \cite{Fisher} for 
$e(m)=(1/2\chi) m^2- (1/12\pi c^2 \chi^3) m^3$. This allows to determine from 
the slopes for $m\to 0$ in Fig.~\protect\ref{ztot}\protect\, the SW velocity
$c/c_0 = 1.18(4)$ 
(upper curve), $c/c_0 = 1.1(10)$ (lower curve) in good agreement 
with the expectations \cite{Ceperley,Sorella,Runge,Sandvik}.

In conclusion we have given convincing evidence that the low energy 
spectrum  of the  2D quantum antiferromagnet is consistent with the
prediction of the (2+1)D-NL$\sigma$M. Unexpected anomalies in the finite 
size spectrum are however detected in the $s=1/2$ HM. The physical 
origin of these size corrections can be understood on the basis of 
linear SW approximation: One of the two gapless excitations present in the
SU(2) model, acquires a mass when an external magnetic field is applied. 
The competition between the massive and the massless mode becomes 
quite strong in the $m\to 0$ limit leading to the observed anomalies.
Making use of the results of SWT, we have been able to obtain
a very accurate estimate for the spin susceptibility.

This extensive discussion of finite size effects in the low energy 
spectrum of the 2D-HM has interesting connections with the finite
temperature behavior of the system. In fact, on the basis of the
mapping to the NL$\sigma$M, the properties of the HM at finite 
temperature $T$ coincide with those at $T=0$ of a HM in 
an infinitely long strip of width $L=1/T$ and periodic boundary conditions. 
By applying the finite size SW analysis to the stripe geometry, 
we then evaluate the uniform magnetization $m$ in an external 
magnetic field $h$ at temperature $T$:
\begin{equation}
m(h,T)=\chi h + {h^2\over 4\pi c^2} + {hT\over 2\pi c^2}\,\ln(1-e^{-h/T}) .
\label{eos} 
\end{equation}
This expression, asymptotically valid for small $h$ and $T$ but arbitrary 
ratio $h/T$, obeys the known scaling form predicted by Fisher \cite{Fisher}.
The anomalous finite size corrections in the energy spectrum 
we have uncovered also suggest the possible presence of a {\sl small}
crossover temperature $T_{\times}$ corresponding to the
maximum in Fig.~\protect\ref{scal}\protect. In the HM, $k_B T_{\times}$ 
can be estimated as few percents of the natural energy scale $J$.
The observable consequences of this possible anomalous behavior of the 
low temperature susceptibility of the 2D HM should be reconsidered
on the basis of this study.

Financial support has been partly provided by INFM through 
PRA-SC (SS) and a grant of cpu time on a Cray C90 at CINECA. 
It is a pleasure to thank A. Angelucci, 
S. Vitiello, L. Guidoni and G.B. Bachelet 
for their help and E. Tosatti for kind hospitality at SISSA (AP).


\begin{references}
\bibitem{Pines} A. Sokol and D. Pines, \prl {\bf 71}, 2813 (1993).
\bibitem{Anderson}	   P.~W.~Anderson in 
				{\it Basic Notions of Condensed Matter Physics}
				 (The Benjamin/Cummings Publishing Company, 
				 Inc., 1984) and references therein
\bibitem{Ziman} H.~Neurberger and T.~Ziman, Phys.~Rev.~B~{\bf 39}, 2608 (1989)
\bibitem{Fisher}	   D.~S.~Fisher, Phys.~Rev.~B~{\bf 39}, 11783 (1989)
\bibitem{Chakravarty}   S.~Chakravarty {\it et al. }, 
				Phys.~Rev.~Lett.~{\bf 60}, 1057 (1988)		
\bibitem{Hald}	   F.~D.~M.~Haldane, J.~Phys.~C~{\bf 14}, 2585 (1981)
\bibitem{Ceperley}	    N.~Trivedi and D.~M.~Ceperley, 
				Phys.~Rev.~B~{\bf 41}, 4552 (1990)
\bibitem{Sorella} 	S.~Sorella and M. Calandra : {\it unpublished}
\bibitem{Runge}         K.~Runge, Phys.~Rev.~B~{\bf 44}, 122252 (1992)
\bibitem{Sandvik}  	A.~W.~Sandvik preprint: cond-mat/9707123 
\bibitem{byline} Data available upon request (c.lavalle@caspur.it).
\bibitem{Parola}              A.~Parola {\it et al. }, 
				Phys.~Rev.~Lett.~{\bf 71}, 4393 (1993)  
\bibitem{Oguchi}       T.~Oguchi, Phys.~Rev.~{\bf 117}, 117 (1960) 
\bibitem{Zhong}               Q.~F.~Zhong and S.~Sorella, 
				Europh.~Lett.~{\bf 21}, 629 (1993)
\bibitem{Franjic}	F.~Franjic, S.~Sorella, 
				Progr.~Theor.~Phys.~{\bf 97}, 399 (1997)
\bibitem{Canali}	   C.~M.~Canali, M.~Wallin, 
				Phys.~Rev.~B~{\bf 48}, 3264 (1993)
\bibitem{Singh}		   R.~R.~P.~Singh, Phys.~Rev.~B~{\bf 47}, 12337 (1993)
\end{references}
\end{document}